\documentclass[tightenlines,superscriptaddress,eqsecnum,floats,nofootinbib,showpacs]
{revtex4}
\usepackage{amssymb}
\usepackage{stmaryrd}
\usepackage{amsmath}
\usepackage{amsfonts}
\usepackage{mathrsfs}
\usepackage{txfonts}

\usepackage{amsmath,amssymb,amsfonts}
\usepackage{graphicx}

\def\be{\begin{equation}}
\def\ee{\end{equation}}
\def\ba{\begin{eqnarray}}
\def\ea{\end{eqnarray}}
\def\nn{\nonumber}

\newcommand{\mubar}{{\bar \mu}} 
\newcommand{\abs}[1]{{\left|{#1}\right|}} 
\newcommand{\Abs}[1]{{\Big|{#1}\Big|}} 
\newcommand{\ket}[1]{\vert{#1}\rangle} 
\newcommand{\bra}[1]{\langle{#1}\vert} 
\newcommand{\kt}{{\tilde{K}}} 
\newcommand{\R}{\mathcal {R}} 
\newcommand{\ints}{{\int_\Sigma}} 
\newcommand{\ct}{\tilde{c}}
\newcommand{\eff}{{eff}} 
\newcommand{\sgn}{\mathrm{sgn}} 
\newcommand{\Tr}{\mathrm{Tr}} 
\newcommand{\grav}{\mathrm{gr}} 
\newcommand{\sca}{\mathrm{sc}} 
\newcommand{\kin}{\mathrm{kin}} 
\newcommand{\hil}{\mathcal{H}} 
\newcommand{\Euc}{H^{E}} 

\begin{document}


\title{Loop quantum modified gravity and its cosmological application}

\author{Xiangdong Zhang\footnote{scxdzhang@scut.edu.cn}}
\affiliation{Department of Physics, South China University of
Technology, GuangZhou 510641, China} \affiliation{Department of
Physics, Beijing Normal University, Beijing 100875, China}

\author{ Yongge Ma\footnote{ mayg@bnu.edu.cn}}
\affiliation{Department of Physics, Beijing Normal University,
Beijing 100875, China}

\begin{abstract}
A general nonperturvative loop quantization procedure for metric modified gravity is
reviewed. As an example, this procedure is applied to
scalar-tensor theories of gravity. The quantum kinematical
framework of these theories is rigorously constructed. Both the
Hamiltonian and master constraint operators are well defined and
proposed to represent quantum dynamics of scalar-tensor theories. As
an application to models, we set up the basic structure of loop quantum Brans-Dicke cosmology. The
effective dynamical equations of loop quantum Brans-Dicke cosmology are also
obtained, which lay a foundation for the phenomenological investigation to possible quantum gravity effects in cosmology.

\pacs{04.60.Pp, 04.50.Kd}
\end{abstract}

\keywords{Modified gravity, Scalar-tensor theories, loop quantum
gravity, connection dynamics}

\maketitle

\section{Introduction}

As a background independent approach to
quantize general relativity (GR), Loop quantum gravity(LQG) has been widely investigated in past 25 years
\cite{Ro04,Th07,As04,Ma07}. Recently, this non-perturbative loop
quantization procedure has been generalized to the metric $f(\R)$
theories,\cite{Zh11,Zh11b} Brsns-Dicke theory \cite{ZM12a} and
scalar-tensor theories\cite{ZM11c}. The fact that this background-independent
quantization method can be successfully extended to those modified
theories of gravity relies on the key observation that these gravity theories
can be reformulated into the connection dynamical formalism with
a compact structure group. The purpose of this paper is to review how to get the connection dynamics of these modified gravity theories and how to quantize these theories by the nonperturbative loop quantization procedure.

In fact, modified
gravity theories have recently received increased attention in
issues related to``dark matter", ``dark energy" and non-trivial
tests on gravity beyond GR. Since 1998, a series
of independent and accurate observations, including type Ia
supernova, cosmic microwave background anisotropy, weak gravity
lens, baryon oscillation, etc, implied that our universe now is
undergoing a period of accelerated expansion\cite{Fr08}. These
results have caused the ``dark energy" problem which is difficult to get a
satisfactory interpretation within the framework of GR. Hence it is
reasonable to consider the other possibility that GR is not a valid
theory of gravity on a galactic or cosmological scale. Besides the well-known $f(\R)$ theories, a
competing relativistic theory of gravity was proposed by
Brans and Dicke in 1961 \cite{BD}, which is apparently compatible
with Mach's principle. To represent a varying ``gravitational
constant", a scalar field is non-minimally coupled to the metric in
Brans-Dicke theory. To interpret the observational results, the
Brans-Dicke theory was generalized by Bergmann \cite{bergmann} and
Wagoner \cite{wagoner} to more general scalar-tensor theories (STT).
Moreover, scalar-tensor theories is also closely related to low
energy effective actions of some string theory (see e.g.
\cite{tayler,maeda,damour}). Since it can
naturally lead to cosmological acceleration in certain models (see
e.g. \cite{Boiss00,Banerjee,Qiang,Boiss11}), the scalar field
in STT of gravity is expected to account for the ``dark energy" problem.
In addition, some models of STT of gravity may even account for the
``dark matter" puzzle \cite{lee,catena,kim}, which was revealed by
the observed rotation curve of galaxy clusters.

There are infinite ways to modify GR. One may suspect which rules we should employed to do the modification. The decisive rule certainly comes from experiments. A large part of the non-trivial tests on gravity theory beyond GR is
closely related to Einstein's equivalence principle (EEP)
\cite{will}. There are many local experiments existed in
solar-system supporting EEP, which implies that gravity should be described by
metric theories. Indeed, STT are a class of representative metric
theories, which have been received most attention. That is why we
us it as an example to demonstrate our general loop quantization
procedure for metric modified gravity theories. It is also worth noting that both the metric $f(\R)$
theories and Palatini $f(\R)$ theories are equivalent
to certain special kinds of STT with the coupling parameter
$\omega(\phi)=0$ and $\omega(\phi)=-\frac32$ respectively\cite{So} Meanwhile
the original Brans-Dicke theory is nothing but the particular case
of constant $\omega$ and vanishing potential of $\phi$.
There are also some other types of modified metric gravity theories
proposed in recent years, such as Horava-Lifshitz theory \cite{Ho} and
critical gravity \cite{LP} etc. Those theories are proposed based on the fact
that GR is nonrenormalizable at perturbative level, while the introduction of
higher order derivative terms might cure this problem. Thus it
is quite interesting to see whether all those kind of metric
theories of gravity could be quantized nonperturbatively.

The following sections of this paper are organized as follows.  We first introduce a general scheme of loop quantization for metric modified gravity in section \ref{section2}. Then we use scalar-tensor theories as an
example to show how our general quantization procedure works. In section
\ref{section3}, we start with Hamiltonian analysis of STT. The coupling parameter of the STT naturally marks off two sectors of the theory. Based on the resulted connection dynamical
formalism of STT, we then quantize the STT by extending the
nonperturbative quantization procedure of LQG in
section \ref{section4} in the way similar to
loop quantum $f(\R)$ gravity \cite{Zh11,Zh11b}. Nevertheless, the
STT that we are considering are a much more general class of metric
theories of gravity than metric $f(\R)$ theories. The Hamiltonian constraint operators in both sectors of the theory can be well defined. To avoid possible quantum anomaly, master constraint program of STT are given in \ref{section5}. For cosmological application of above quantum gravity theories,
we set up the basic structure of loop quantum Brans-Dicke cosmology and get its effective equations of motion in section
\ref{section6}. Finally some
concluding remarks are given in the last section.
Throughout the paper, we use Greek alphabet for spacetime indices,
Latin alphabet a,b,c,..., for spatial indices, and i,j,k,..., for
internal indices.

\section{General scheme\label{section2}}
In this section, we will first outline the general scheme of loop
quantization for metric modified gravity\cite{Ma12a}. Especially, we are
mainly focus on 4-dimensional metric theories of gravity which is
consistent with Einstein's equivalent principle. The prerequisite is
that the theory which is under consideration should have a well-defined geometrical dynamics,
which means a Hamiltonian formalism with 3-metric $h_{ab}$ as one of
configuration variables, and in addition the constraint algebra of this
theory is first-class (perhaps after solving some second-class
constraints). Without loss of generality, we can assume that the
classical phase space of this theory consists of conjugate pairs
$(h_{ab},p^{ab} )$ and $(\phi_B,\pi^B)$, where $\phi_B$ could be a
scalar, vector, tensor or spinor field. Then the quantization scheme has the following recipe.

(i) To
obtain a connection dynamical formalism, we first define a quantity
$\kt_{ab}$ via \ba \kt_{ab} =
\frac{2\kappa}{\sqrt{h}}\left(p_{ab}-\frac12ph_{ab}\right).\ea Then
we enlarge the phase space by transforming to the triad formulation
as \ba(h_{ab},p^{ab} ) \Rightarrow  (E^a_i \equiv\sqrt{h}h_{ab}
e^a_i , \kt^i_a\equiv\kt_{ab}e^b_i ).\ea Now we make a canonical
transformation to connection formulation as: \ba (E^a_j ,
\kt^j_a)\Rightarrow (E^a_j ,A^j_a\equiv\Gamma^j_a
+\gamma\kt^j_a),\ea and due to symmetric property of $p_{ab}$ we
have $ \kt_{a[i}E^a_{j]} =0$. This will give us the Gaussian
constraint, $\mathcal {D}_aE^a_i\equiv \partial_aE^a_i +
\epsilon_{ijk}A^j_aE^a_k=0$. Then it is straightforward to write all
the constraints in terms of the new variables. (ii) For loop
quantization, we first represent the fields $(\phi_B,\pi^B)$ via
polymer-like representation, together with the LQG representation
for the holonomy-flux algebra. Then the kinematical Hilbert space
can be read as $\hil_\kin := \hil^\grav_\kin \otimes\hil^\phi_\kin$.
All the basic operators and geometrical operators could be well
defined in this Hilbert space. We can solve the Gaussian and
diffeomorphism constraints as in standard LQG. Then we would get the
gauge and diffeomorphism invariant Hilbert spaceas:
$\hil_\kin\dashrightarrow \hil_G \dashrightarrow\hil_{Diff} $. In
order to implement quantum dynamics, the Hamiltonian constraint
operator may first be constructed at least in $\hil_G$, although it
usually could not be well defined in $\hil_{Diff}$. Then the master
constraint operator can be constructed in $\hil_{Diff}$ by using the
structure of the Hamiltonian operator. (iii) One may try to
understand the physical Hilbert space by the direct integral
decomposition of $\hil_{Diff}$ with respect to the spectrum of the master constraint
operator. (iv) One may also do certain semiclassical analysis in
order to confirm the classical limits of the Hamiltonian and master
constraint operators as well as the constraint algebra. The low
energy physics is also expected in the analysis. (v) Finally, to
complement above canonical approach, we can also try the covariant
path integral (spinfoam) quantization.

It should be noted that the last three steps are still open issues in the loop quantization of GR. Thus in the following sections, we will take scalar-tensor theories as an example to carry out the steps (i) and (ii) in the above scheme of loop quantization for metric modified gravity.

\section{Hamiltonian analysis of scalar-tensor theories\label{section3}}

The most general action of STT reads \ba
S(g)=\frac{1}{2\kappa}\int_\Sigma
d^4x\sqrt{-g}\left[\phi\R-\frac{\omega(\phi)}{\phi}(\partial_\mu\phi)\partial^\mu\phi-2\xi(\phi)\right]\label{action}
\ea where $\kappa=8\pi G$, $\R$ denotes the scalar curvature of
spacetime metric $g_{\mu\nu}$, and the coupling parameter $\omega(\phi)$
and potential $\xi(\phi)$ can be arbitrary functions of scalar field
$\phi$. Variations of the action (\ref{action}) with respect to
$g_{ab}$ and $\phi$ respectively give us equations of motion
\ba &&\phi
G_{\mu\nu}=\nabla_\mu\nabla_\nu\phi-g_{\mu\nu}\Box\phi
+\frac{\omega(\phi)}{\phi}[(\partial_\mu\phi)\partial_\nu\phi-\frac12g_{\mu\nu}(\nabla\phi)^2]-g_{\mu\nu}\xi(\phi),\label{01}
\\
&&\R+\frac{2\omega(\phi)}{\phi}\Box\phi-\frac{\omega(\phi)}{\phi^2}(\partial_\mu\phi)\partial^\mu\phi
+\frac{\omega'(\phi)}{\phi}(\partial_\mu\phi)\partial^\mu\phi-2\xi'(\phi)=0,\label{02}\ea
where a prime over a letter denotes a derivative with respect to the
argument,  $\nabla_\mu$ is the covariant derivative compatible with
$g_{\mu\nu}$ and $\Box\equiv g^{\mu\nu}\nabla_\mu\nabla_\nu$. By
doing 3+1 decomposition of the spacetime, the four-dimensional (4d)
scalar curvature can be expressed as \ba \mathcal
{R}=K_{ab}K^{ab}-K^2+R+\frac{2}{\sqrt{-g}}\partial_\mu(\sqrt{-g}n^\mu
K)-\frac{2}{N\sqrt{h}}\partial_a
(\sqrt{h}h^{ab}\partial_bN)\label{03} \ea where $K_{ab}$ is the
extrinsic curvature of a spatial hypersurface $\Sigma$, $K\equiv
K_{ab}h^{ab}$, $R$ denotes the scalar curvature of the 3-metric
$h_{ab}$ induced on $\Sigma$, $n^\mu$ is the unit normal of $\Sigma$
and $N$ is the lapse function. By Legendre transformation, the
momenta conjugate to the dynamical variables $h_{ab}$ and $\phi$ are
defined respectively as
\ba p^{ab}&=&\frac{\partial\mathcal
{L}}{\partial\dot{h}_{ab}}=\frac{\sqrt{h}}{2\kappa}[\phi(K^{ab}-Kh^{ab})-\frac{h^{ab}}{N}(\dot{\phi}-N^c\partial_c\phi)], \label{04}\\
\pi&=&\frac{\partial\mathcal
{L}}{\partial\dot{\phi}}=-\frac{\sqrt{h}}{\kappa}(K-\frac{\omega(\phi)}{N\phi}(\dot{\phi}-N^c\partial_c\phi)),\label{pi}
\ea where $N^c$ is the shift vector. The combination of the trace of Eq.
(\ref{04}) and Eq. (\ref{pi}) gives \ba
(3+2\omega(\phi))(\dot{\phi}-N^a\partial_a\phi)=\frac{2\kappa
N}{\sqrt{h}}(\phi\pi-p).\label{Sconstraint} \ea It is easy to see
from Eq. (\ref{Sconstraint}) that one extra constraint
$S=p-\phi\pi=0$ emerges when $\omega(\phi)=-\frac32$. Hence it is
natural to mark off two sectors of the theories by
$\omega(\phi)\neq-\frac32$ and $\omega(\phi)=-\frac32$.

\subsection{Sector of $\omega(\phi)\neq -3/2$ }

In the case of $\omega(\phi)\neq -3/2$, the Hamiltonian of STT can
be derived as a liner combination of constraints as
\ba H_{total}=\int_\Sigma d^3x(N^aV_a+NH),\label{htotal} \ea
where
the smeared diffeomorphism and Hamiltonian constraints read
respectively
\ba V(\overrightarrow{N})&=&\int_\Sigma d^3xN^aV_a =\int_\Sigma
d^3xN^a\left(-2D^b(p_{ab})+\pi\partial_a\phi\right),\label{dc}\\
H(N)&=&\int_\Sigma d^3xNH \nn\\
&=&\int_\Sigma
d^3xN\left[\frac{2\kappa}{\sqrt{h}}\left(\frac{p_{ab}p^{ab}-\frac12p^2}{\phi}+\frac{(p-\phi\pi)^2}{2\phi(3+2\omega)}\right)
+\frac{\sqrt{h}}{2\kappa}\left(-\phi
R+\frac{\omega(\phi)}{\phi}(D_a\phi)
D^a\phi+2D_aD^a\phi+2\xi(\phi)\right)\right].\label{hc}\nn\\
\ea
By the symplectic structure
\ba
\{h_{ab}(x),p^{cd}(y)\}&=&\delta^{(c}_a\delta^{d)}_b\delta^3(x,y),\nn\\
\{\phi(x),\pi(y)\}&=&\delta^3(x,y), \label{poission}\ea lengthy but
straightforward calculations show that the constraints (\ref{dc})
and (\ref{hc}) comprise a first-class system similar to GR as:
\ba \{V(\overrightarrow{N}),V(\overrightarrow{N}^\prime)\}&=& V([\overrightarrow{N},\overrightarrow{N}^\prime]), \nn\\
\{H(M),V(\overrightarrow{N})\}&=&- H(\mathcal
{L}_{\overrightarrow{N}}M), \nn\\
\{H(N),H(M)\}&=& V(ND^aM-MD^aN). \ea  We can show that the above
Hamiltonian formalism of STT is equivalent to their Lagrangian
formalism\cite{ZM11c}.

To obtain the connection dynamical formalism of the STT, following
the general scheme mentioned in last section, we define
\ba\tilde{K}^{ab}=\phi
K^{ab}+\frac{h^{ab}}{2N}(\dot{\phi}-N^c\partial_c\phi)=\phi
K^{ab}+\frac{h^{ab}}{(3+2\omega)\sqrt{h}}(\phi\pi-p).\ea
Then we can introduce new canonical pairs
$(E^a_i\equiv\sqrt{h}e^a_i, \tilde{K}_a^i\equiv\kt_{ab}e^b_i)$, where
$e^a_i$ is the triad such that $h_{ab}e^a_ie^b_j=\delta_{ij}$.
Now
the symplectic structure (\ref{poission}) can be got from the
following Poisson brackets: \ba
\{E^a_j(x),E^b_k(y)\}=\{\tilde{K}_a^j(x),\tilde{K}_b^k(y)\}=0,\nn\\
\{\tilde{K}^j_a(x),E_k^b(y)\}=\kappa\delta^b_a\delta^j_k\delta(x,y).
\ea Note that since we have $\tilde{K}^{ab}=\tilde{K}^{ba}$, there
have an additional constraint: \be
G_{jk}\equiv\tilde{K}_{a[j}E^a_{k]}=0. \label{gaussian}\ee
Now we
can make a second canonical transformation via defining: \be
A^i_a=\Gamma^i_a+\gamma\tilde{K}^i_a, \label{newvaribles}\ee
where
$\Gamma^i_a$ is the spin connection determined by $E^a_i$, and
$\gamma$ is a nonzero real number. It is easy to see that our new
variable $A^j_a$ coincides with the Ashtekar-Barbero connection
\cite{As86,Ba} in the special case $\phi=1$. The Poisson brackets
among the new variables read: \ba
\{A^j_a(x),E_k^b(y)\}&=&\gamma\kappa\delta^b_a\delta^j_k\delta(x,y),\nn\\
\{A_a^i(x),A_b^j(y)\}&=&0,\quad \{E_j^a(x),E_k^b(y)\}=0. \ea
Now
the phase space of the STT consists of conjugate pairs
$(A_a^i,E^b_j)$ and $(\phi,\pi)$. Combining Eq.(\ref{gaussian}) with
the compatibility condition: \ba
\partial_aE^a_i+\epsilon_{ijk}\Gamma^j_aE^{ak}=0,
\ea the standard Gaussian constraint can be obtained as
\ba \mathcal
{G}_i=\mathscr{D}_aE^a_i\equiv\partial_aE^a_i+\epsilon_{ijk}A^j_aE^{ak},
\label{GC}\ea
which justifies $A^i_a$ as an $su(2)$-connection. It
is worth noting that, had we let $\gamma=\pm i$, the
(anti-)self-dual complex connection formalism would be obtained. The
original vector and Hamiltonian constraints can be
written in terms of new variables up to Gaussian constraint respectively as \ba
V_a &=&\frac1{\kappa\gamma} F^i_{ab}E^b_i+\pi\partial_a\phi,
\label{diff}\ea
\ba H
&=&\frac{\phi}{2\kappa}\left[F^j_{ab}-(\gamma^2+\frac{1}{\phi^2})\varepsilon_{jmn}\tilde{K}^m_a\tilde{K}^n_b\right]
\frac{\varepsilon_{jkl}
E^a_kE^b_l}{\sqrt{h}}\nn\\
&+&\frac{\kappa}{(3+2\omega(\phi))}\left(\frac{(\tilde{K}^i_aE^a_i)^2}{\kappa^2\phi\sqrt{h}}+
2\frac{(\tilde{K}^i_aE^a_i)\pi}{\kappa\sqrt{h}}+\frac{\pi^2\phi}{\sqrt{h}}\right) \nn\\
&+&\frac{1}{\kappa}\left[\frac{\omega(\phi)}{2\phi}\sqrt{h}(D_a\phi)
D^a\phi+\sqrt{h}D_aD^a\phi+\sqrt{h}\xi(\phi)\right],\label{hamilton}
\ea where
$F^i_{ab}\equiv2\partial_{[a}A^i_{b]}+\epsilon^i_{\ kl}A_a^kA_b^l$ is
the curvature of $A_a^i$. The total Hamiltonian can be expressed as
a linear combination
\ba H_{total}=\ints\Lambda^i\mathcal
{G}_i+N^aV_a+NH.\ea
It is easy to check that the smeared Gaussian
constraint, $\mathcal {G}(\Lambda):=\int_\Sigma
d^3x\Lambda^i(x)\mathcal {G}_i(x)$, generates $SU(2)$ gauge
transformations on the phase space, while the smeared constraint,
$\mathcal {V}(\overrightarrow{N}):=\int_\Sigma
d^3xN^a(V_a-A_a^i\mathcal {G}_i)$, generates spatial diffeomorphism
transformations on the phase space. Together with the smeared
Hamiltonian constraint $H(N)=\int_\Sigma d^3xNH$, we can show that
the constraints algebra has the following form: \ba \{\mathcal
{G}(\Lambda),\mathcal {G}(\Lambda^\prime)\}&=&\kappa\mathcal
{G}([\Lambda,\Lambda^\prime]),\label{eqsA} \\
\{\mathcal
{G}(\Lambda),\mathcal{V}(\overrightarrow{N})\}&=&-\mathcal{G}(\mathcal
{L}_{\overrightarrow{N}}\Lambda,),\\
\{\mathcal {G}(\Lambda),H(N)\}&=&0,\\
\{\mathcal {V}(\overrightarrow{N}),\mathcal
{V}(\overrightarrow{N}^\prime)\}&=&\mathcal
{V}([\overrightarrow{N},\overrightarrow{N}^\prime]), \\
\{\mathcal {V}(\overrightarrow{N}),H(M)\}&=& H(\mathcal
{L}_{\overrightarrow{N}}M),\label{eqsE}\\
\{H(N),H(M)\}&=&\mathcal {V}(ND^aM-MD^aN)\nn\\
&+&\mathcal
{G}\left((N\partial_aM-M\partial_aN)h^{ab}A_b\right)\nn\\
&-&\frac{[E^aD_aN,E^bD_bM]^i}{\kappa h}\mathcal {G}_i\nn\\
&-&\gamma^2\frac{[E^aD_a(\phi N),E^bD_b(\phi M)]^i}{\kappa
h}\mathcal {G}_i.\label{eqsb}\ea Here Eqs.(\ref{eqsA}-\ref{eqsE})
can be understand by the geometrical interpretations of $\mathcal
{G}(\Lambda)$ and $\mathcal {V}(\overrightarrow{N})$. The detail
calculation on the Poisson bracket (\ref{eqsb}) between the two
smeared Hamiltonian constraints can be seen in the Appendix of
\cite{ZM11c}. Hence the constraints are of first class. Furthermore,
the constraint algebra of GR can be recovered for the special case
when $\phi=1$. To summarize, the STT of gravity in the sector
$\omega(\phi)\neq -3/2$ have already been cast into the
$su(2)$-connection dynamical formalism. The resulted Hamiltonian
structure is quite similar to metric $f(R)$ theories\cite{Zh11b}.

\subsection{Sector of $\omega(\phi)= -3/2$ }

In the  sector of $\omega(\phi)= -3/2$, Eq. (\ref{Sconstraint})
implies that there is an extra primary constraint $S=0$, which we call
``conformal" constraint. Hence, as pointed out in
Ref.\cite{olmo}, the total Hamiltonian in this case can be
expressed as a liner combination of constraints as \ba
H_{total}=\int_\Sigma d^3x(N^aV_a+NH+\lambda S),\label{htotal1} \ea
where the smeared diffeomorphism constraint $V(\overrightarrow{N})$ is as same
as (\ref{dc}), while the Hamiltonian and conformal constraints read
respectively \ba
H(N)&=&\int_\Sigma d^3xNH \nn\\
&=&\int_\Sigma
d^3xN\left[\frac{2\kappa}{\sqrt{h}}\left(\frac{p_{ab}p^{ab}-\frac12p^2}{\phi}\right)
+\frac{\sqrt{h}}{2\kappa}(-\phi R-\frac{3}{2\phi}(D_a\phi)
D^a\phi+2D_aD^a\phi+2\xi(\phi))\right],\label{hc1}\\
S(\lambda)&=&\int_\Sigma d^3x\lambda S=\int_\Sigma
d^3x\lambda(p-\phi\pi).\label{sc}\ea
With the help of symplectic
structure (\ref{poission}), straightforward calculations show that \ba
\{H(M),V(\overrightarrow{N})\}&=&- H(\mathcal
{L}_{\overrightarrow{N}}M),\quad
\{S(\lambda),V(\overrightarrow{N})\}=- S(\mathcal
{L}_{\overrightarrow{N}}\lambda),\label{VHS}\\
\{H(N),H(M)\}&=& V(ND^aM-MD^aN)+ S(\frac{D_a\phi}{\phi}(ND^aM-MD^aN)),\label{HH}\\
\{S(\lambda),H(M)\}&=& H(\frac{\lambda M}{2})+\ints
N\lambda\sqrt{h}(-2\xi(\phi)+\phi\xi'(\phi)).\label{Sc} \ea
The
Poisson bracket (\ref{Sc}) implies that, we have to impose a
secondary constraint in order to maintain the constraints $S$ and $H$ in the
time evolution as
\ba -2\xi(\phi)+\phi\xi'(\phi)=0.
\label{equationofV}\ea It is clear that this constraint is
second-class, and hence one has to solve it. Here we consider the
vacuum case where the solutions of Eq. (\ref{equationofV}) are either
$\xi(\phi)=0$ or $\xi(\phi)=C\phi^2$, where $C$ is certain
dimensional constant. Thus the consistency condition strongly
restricted the feasible STT in this sector to only two theories. As
pointed out in Ref.\cite{So}, for these two theories, the action
(\ref{action}) is invariant under the following conformal
transformation:
\ba g_{\mu\nu}\rightarrow e^\lambda g_{\mu\nu},\quad
\phi\rightarrow e^{-\lambda}\phi.\label{conformalt} \ea
Thus,
besides diffeomorphism invariance, those two theories are also
conformally invariant. Now in the resulted Hamiltonian formalism the
constraints $(V,H,S)$ form a first-class system. The following
transformations on the phase space are generated by the conformal
constraint
\ba &&\{h_{ab},S(\lambda)\}=\lambda h_{ab},\quad
\{P^{ab},S(\lambda)\}=-\lambda P^{ab}, \\
&&\{\phi,S(\lambda)\}=-\lambda \phi,\quad \{\pi,S(\lambda)\}=\lambda
\pi. \ea
It is clear that the above transformations coincide with
those of spacetime conformal transformations (\ref{conformalt}).
Hence all constraints have clear physical meaning. Now the physical
degrees of freedom of this special kind of STT are equal to those of
GR because of the extra conformal constraint (\ref{sc}).

The connection-dynamical formalism for STT in this sector can also
be obtained by the canonical transformations to the new variables
(\ref{newvaribles}). Then the total Hamiltonian can be expressed
again as a linear combination
\ba H_{total}=\ints\Lambda^i\mathcal
{G}_i+N^aV_a+NH+\lambda S,\ea
where the Gaussian and diffeomorphism
constraints keep the same form as Eqs. (\ref{GC}) and (\ref{diff}),
while the Hamiltonian and the conformal constraints read
respectively
\ba  H
&=&\frac{\phi}{2\kappa}\left[F^j_{ab}-(\gamma^2+\frac{1}{\phi^2})\varepsilon_{jmn}\tilde{K}^m_a\tilde{K}^n_b\right]
\frac{\varepsilon_{jkl}
E^a_kE^b_l}{\sqrt{h}}\nn\\
&+&\frac1\kappa\Big[-\frac{3}{4\phi}\sqrt{h}(D_a\phi)
D^a\phi+\sqrt{h}D_aD^a\phi+\sqrt{h}\xi(\phi)\Big],\label{hamilton1}\\
S&=&\frac{1}{\kappa}\tilde{K}^i_aE^a_i-\pi\phi
\label{conformalc}.\ea
Now the constraints algebra in the connection
formalism is closed as \ba
\{\mathcal {G}(\Lambda),H(N)\}&=&0,\\
\{\mathcal {G}(\Lambda),S(\lambda)\}&=&0,\\
\{S(\lambda),H(M)\}&=& H(\frac{\lambda M}{2}),\\
\{H(N),H(M)\}&=&\Big[\mathcal {V}(ND^aM-MD^aN)\nn\\
&+&S(\frac{D_a\phi}{\phi}(ND^aM-MD^aN)) \nn\\
&+&\mathcal
{G}\left((N\partial_aM-M\partial_aN)h^{ab}A_b\right)\nn\\
&-&\frac{[E^aD_aN,E^bD_bM]^i}{\kappa h}\mathcal {G}_i\nn\\
&-&\gamma^2\frac{[E^aD_a(\phi N),E^bD_b(\phi M)]^i}{\kappa
h}\mathcal {G}_i\Big].\ea
It is easy to see that the Poisson brackets
among the other constraints weakly equal to zero. Hence we have cast
STT of gravity with $\omega(\phi)\neq -3/2$ into the
$su(2)$-connection dynamical formalism.

\section{Loop quantization of scalar-tensor theories\label{section4}}

Based on the connection dynamical formalism obtained in last section,
the nonperturbative loop quantization procedure can be naturally
extended to the STT. The kinematical structure of STT is as same as
that of LQG coupled with a scalar field and $f(\R)$ theories \cite{Zh11,Zh11b}. The kinematical
Hilbert space of the system is a direct product of the Hilbert
space of geometry part and that of scalar field,
$\hil_\kin:=\hil^\grav_\kin\otimes \hil^\sca_\kin$. An orthonormal
basis of this Hilbert space is the so called spin-scalar-network
basis, $T_{\alpha,X}(A,\phi)\equiv T_{\alpha}(A)\otimes T_{X}(\phi)$,
over some graph $\alpha\cup X\subset\Sigma$, where $\alpha$ and $X$
consist of finite number of curves and points in $\Sigma$
respectively. The basic operators of quantum STT are the quantum
analogue of holonomies $h_e(A)=\mathcal {P}\exp\int_eA_a$ of a
connection along edges $e\subset\Sigma$, densitized triads
$E(S,f):=\int_S\epsilon_{abc}E^a_if^i$ smeared over
2-surfaces, point holonomies
$U_\lambda=\exp(i\lambda\phi(x))$\cite{As03}, and scalar momenta
$\pi(R):=\int_R d^3x\pi(x)$ smeared on 3-dimensional regions. It is
worth noting that the spatial geometric operator, such as the
area\cite{Ro95} , the volume\cite{As97} and the
length\cite{Th98,Ma10} operators, are still valid in
$\hil^\grav_\kin$ of quantum STT. As in LQG, it is natural
to promote the Gaussian constraint $\mathcal {G}(\Lambda)$ as a
well-defined operator\cite{Th07,Ma07}. Then it's kernel is the
internal gauge invariant Hilbert space $\mathcal {H}_G$ with
gauge invariant spin-scalar-network basis. Since the diffeomorphisms
of $\Sigma$ act covariantly on the cylindrical functions in
$\mathcal {H}_G$, the group averaging technique can be employed to
solve the diffeomorphism constraint \cite{As04,Ma07,Zh11b}.
Hence the desired diffeomorphism and gauge invariant Hilbert space
$\mathcal {H}_{Diff}$ for the STT  can also be
obtained\cite{Zh11b,ZM11c}.

\subsection{Sector of $\omega(\phi)\neq -3/2$ }

While the kinematical framework of LQG and polymer-like scalar field
have been straight-forwardly extended to the STT, the nontrivial
task in the case of $\omega(\phi)\neq -3/2$ is to implement the
Hamiltonian constraint (\ref{hamilton}) at quantum level. In order
to compare the Hamiltonian constraint of STT with that of metric
$f(\R)$ theories in connection formalism, we write Eq.
(\ref{hamilton}) as $H(N)=\sum^8_{i=1}H_i$ in regular order. It is
easy to see that the terms $H_1,H_2,H_7,H_8$ just keep the same form
as those in $f(\R)$ theories (see Eq.(39) in Ref.\cite{Zh11b}), and
the $H_3,H_4,H_5$ terms are also similar to the corresponding terms
in $f(\R)$ theories. Here the differences are only reflected by the
coefficients as certain functions of $\phi$. Now we come to the
completely new term, $H_6=\int_\Sigma
d^3xN\frac{\omega(\phi)}{2\phi}\sqrt{h}(D_a\phi) D^a\phi $. This
term is somehow like the kinetic term of a Klein-Gordon field which
was dealt with in Ref.\cite{Ma06}. We can introduce the well-defined
operators $\phi,\phi^{-1}$ as in Ref. \cite{Zh11b}. It is reasonable
to believe that function $\omega(\phi)$ can also be quantized
\cite{Zh11b}. By the same regularization techniques as in
Refs.\cite{Ma06,Zh11b}, we triangulate $\Sigma$ in adaptation to
some graph $\alpha$ underling a cylindrical function in $\hil_\kin$
and reexpress connections by holonomies. The corresponding regulated
operator acts on a basis vector $T_{\alpha,X}$ over some graph
$\alpha\cup X$ as \ba \hat{H}^\varepsilon_6\cdot T_{\alpha,X}
&=&\lim_{\epsilon\rightarrow 0}\frac{2^{17}N(v)\hat{\omega}(\phi)
}{3^6\gamma^4(i\lambda_0)^2(i\hbar)^4\kappa}\hat{\phi}^{-1}(v)
\chi_\epsilon(v-v')
\nn\\
&\times&\sum_{v\in\alpha(v)}\frac{1}{E(v)}\sum_{v(\Delta)=v}\epsilon(s_L
s_M s_N)\epsilon^{LMN}\hat{U}^{-1}_{\lambda_0}(\phi(s_{s_L(\Delta_{v})}))\nn\\
&\times&
[\hat{U}_{\lambda_0}(\phi(t_{s_L(\Delta_{v})}))-\hat{U}_{\lambda_0}(\phi(s_{s_L(\Delta_{v})}))]\nn\\
&\times&\Tr(\tau_i\hat{h}_{s_M(\Delta_{v})}[\hat{h}^{-1}_{s_M(\Delta_{v})},(\hat{V}_{U^\epsilon_{v}})^{3/4}]
\hat{h}_{s_N(\Delta_{v})}[\hat{h}^{-1}_{s_N(\Delta_{v})},(\hat{V}_{U^\epsilon_{v}})^{3/4}]) \nn\\
&\times&\sum_{v'\in\alpha(v)}\frac{1}{E(v')}\sum_{v(\Delta')=v'}\epsilon(s_I
s_J s_K)\epsilon^{IJK}\hat{U}^{-1}_{\lambda_0}(\phi(s_{s_I(\Delta_{v'})}))\nn\\
&\times&
[\hat{U}_{\lambda_0}(\phi(t_{s_I(\Delta_{v'})}))-\hat{U}_{\lambda_0}(\phi(s_{s_I(\Delta_{v'})}))]\nn\\
&\times&\Tr(\tau_i\hat{h}_{s_J(\Delta_{v'})}[\hat{h}^{-1}_{s_J(\Delta_{v'})},(\hat{V}_{U^\epsilon_{v'}})^{3/4}]
\hat{h}_{s_K(\Delta_{v'})}[\hat{h}^{-1}_{s_K(\Delta_{v'})},(\hat{V}_{U^\epsilon_{v'}})^{3/4}])\cdot
T_{\alpha,X}. \label{H6}\ea
We refer to \cite{Zh11b} for the meaning
of notations in Eq.(\ref{H6}). It is easy to see that the action of
$\hat{H}^\varepsilon_6$ on $ T_{\alpha,X}$ is graph changing. It
adds a finite number of vertices at $t(s_I(v))=\varepsilon$ for
edges $e_I(t)$ starting from each high-valent vertex of $\alpha$. As
a result, the family of operators $\hat{H}^\varepsilon_6(N)$ fails
to be weakly convergent when $\varepsilon\rightarrow 0$. However,
due to the diffeomorphism covariant properties of the triangulation,
the limit operator can be well defined via the so-called uniform
Rovelli-Smolin topology induced by diffeomorphism-invariant states
$\Phi_{Diff}$ as:
\ba \Phi_{Diff}(\hat{H}_6\cdot
T_{\alpha,X})=\lim_{\varepsilon\rightarrow
0}(\Phi_{Diff}|\hat{H}^\varepsilon_{6}|T_{\alpha,X}\rangle. \ea
It
is obviously that the limit is independent of $\varepsilon$. Hence
both the regulators $\epsilon$ and $\varepsilon$ can be removed. We
then have
\ba \hat{H}_6\cdot T_{\alpha,X} &=&\sum_{v\in
V(\alpha)}\frac{2^{17}N(v)\hat{\omega}(\phi) }{3^6\gamma^4(i\lambda_0)^2(i\hbar)^4\kappa E^2(v)}\hat{\phi}^{-1}(v)\nn\\
&\times&\sum_{v(\Delta)=v(\Delta')=v}\epsilon(s_L
s_M s_N)\epsilon^{LMN}\hat{U}^{-1}_{\lambda_0}(\phi(s_{s_L(\Delta)}))\nn\\
&\times&
[\hat{U}_{\lambda_0}(\phi(t_{s_L(\Delta)}))-\hat{U}_{\lambda_0}(\phi(s_{s_L(\Delta)}))]\nn\\
&\times&\Tr(\tau_i\hat{h}_{s_M(\Delta)}[\hat{h}^{-1}_{s_M(\Delta)},(\hat{V}_v)^{3/4}]
\hat{h}_{s_N(\Delta)}[\hat{h}^{-1}_{s_N(\Delta)},(\hat{V}_v)^{3/4}]) \nn\\
&\times&\epsilon(s_I
s_J s_K)\epsilon^{IJK}\hat{U}^{-1}_{\lambda_0}(\phi(s_{s_I(\Delta')}))\nn\\
&\times&
[\hat{U}_{\lambda_0}(\phi(t_{s_I(\Delta')}))-\hat{U}_{\lambda_0}(\phi(s_{s_I(\Delta')}))]\nn\\
&\times&\Tr(\tau_i\hat{h}_{s_J(\Delta')}[\hat{h}^{-1}_{s_J(\Delta')},(\hat{V}_{v})^{3/4}]
\hat{h}_{s_K(\Delta')}[\hat{h}^{-1}_{s_K(\Delta')},(\hat{V}_{v})^{3/4}])\cdot
T_{\alpha,X} . \ea In order to simplify the expression, we introduce
$f(\phi)=\frac{1}{3+2\omega(\phi)}$ for the other terms containing
it in $H(N)$, which can also be promoted to a well-defined operator
$\hat{f}(\phi)$. Hence, the terms $H_3,H_4$ and $H_5$ can be
quantized as \ba \hat{H}_3\cdot T_{\alpha,X} &=&
\sum_{v\in V(\alpha)}\frac{4N(v)\hat{f}(\phi(v))}{\gamma^3(i\hbar)^2\kappa}\hat{\phi}^{-1}(v)\nn\\
&\times&
[\hat{\Euc}(1),\sqrt{\hat{V}_v}] [\hat{\Euc}(1),\sqrt{\hat{V}_{v}}]\cdot T_{\alpha,X}, \nn\\
\ea

\ba \hat{H}_4\cdot T_{\alpha,X} &=&-\sum_{v\in V(\alpha)\cap
X}\frac{2^{20}N(v)\hat{f}(\phi(v))}{3^5\gamma^6(i\hbar)^6
E^2(v)}\hat{\pi}(v)
\nn\\
&\times&\sum_{v(\Delta)=v(\Delta')=v}\Tr(\tau_i\hat{h}_{s_L(\Delta)}[\hat{h}^{-1}_{s_L(\Delta)},\hat{\kt}])\nn\\
&\times&\epsilon(s_L s_M s_N)\epsilon^{LMN}\nn\\
&\times&\Tr(\tau_i\hat{h}_{s_M(\Delta)}[\hat{h}^{-1}_{s_M(\Delta)},(\hat{V}_{v})^{3/4}]
\hat{h}_{s_N(\Delta)}[\hat{h}^{-1}_{s_N(\Delta)},(\hat{V}_{v})^{3/4}]) \nn\\
&\times&\epsilon(s_I s_J s_K)\epsilon^{IJK}\nn\\
&\times&\Tr(\hat{h}_{s_I(\Delta')}[\hat{h}^{-1}_{s_I(\Delta')},(\hat{V}_{v})^{1/2}]
\hat{h}_{s_J(\Delta')}[\hat{h}^{-1}_{s_J(\Delta')},(\hat{V}_{v})^{1/2}] \nn\\
&\times&\hat{h}_{s_K(\Delta')}[\hat{h}^{-1}_{s_K(\Delta')},(\hat{V}_{v})^{1/2}])\cdot
T_{\alpha,X}, \ea

\ba \hat{H}_5\cdot T_{\alpha,X} &=&\sum_{v\in V(\alpha)\cap
X}\frac{2^{18}\kappa N(v)\hat{f}(\phi(v))}{3^4\gamma^6(i\hbar)^6
E^2(v)}
\hat{\phi}(v)\hat{\pi}(v)\hat{\pi}(v) \nn\\
&\times&\sum_{v(\Delta)=v(\Delta')=v}\epsilon(s_I s_J
s_K)\epsilon^{IJK}\nn\\
&\times&\Tr(\hat{h}_{s_I(\Delta)}
[\hat{h}^{-1}_{s_I(\Delta)},(\hat{V}_{v})^{1/2}]
\hat{h}_{s_J(\Delta)}[\hat{h}^{-1}_{s_J(\Delta)},(\hat{V}_{v})^{1/2}] \nn\\
&\times&\hat{h}_{s_K(\Delta)}[\hat{h}^{-1}_{s_K(\Delta)},(\hat{V}_{v})^{1/2}]) \nn\\
&\times&\epsilon(s_L s_M s_N)\epsilon^{LMN}\nn\\
&\times&\Tr(\hat{h}_{s_L(\Delta')}[\hat{h}^{-1}_{s_L(\Delta')},(\hat{V}_{v})^{1/2}]
\hat{h}_{s_M(\Delta')}[\hat{h}^{-1}_{s_M(\Delta')},(\hat{V}_{v})^{1/2}] \nn\\
&\times&\hat{h}_{s_N(\Delta')}[\hat{h}^{-1}_{s_N(\Delta')},(\hat{V}_{v})^{1/2}])\cdot
T_{\alpha,X}. \ea
While the $H_7$ and $H_8$ terms keep the same form
as in $f(\R)$ theory, which read respectively
\ba \hat{H}_7\cdot
T_{\alpha,X} &=&\sum_{v\in
V(\alpha)}\frac{2^{7}N(v)}{3\gamma^2i\lambda_0(i\hbar)^2\kappa E(v)}\nn\\
&\times&\sum_{e(0)=v}X^i_e\sum_{v(\Delta)=v}\epsilon(s_I s_J
s_K)\epsilon^{IJK}\nn\\
&\times&\hat{U}^{-1}_{\lambda_0}(\phi(s_{s_I(\Delta)}))
[\hat{U}_{\lambda_0}(\phi(t_{s_I(\Delta)}))-\hat{U}_{\lambda_0}(\phi(s_{s_I(\Delta)}))]\nn\\
&\times&\Tr(\tau_i\hat{h}_{s_J(\Delta)})[\hat{h}^{-1}_{s_J(\Delta)},(\hat{V}_{v})^{1/2}]\nn\\
&\times&
\hat{h}_{s_K(\Delta)}[\hat{h}^{-1}_{s_K(\Delta)},(\hat{V}_{v})^{1/2}])\cdot
T_{\alpha,X}, \ea
\ba \hat{H}_8\cdot T_{\alpha,X} &=&\frac1\kappa\sum_{v\in
V(\alpha)}N(v)\hat{\xi}(\phi(v))\hat{V_v}\cdot T_{\alpha,X}. \ea
Here the action of the volume operator $\hat{V}$ on a spin-network
basis vector $T_\alpha(A)$ over a graph $\alpha$ can be factorized
as \ba \hat{V}\cdot T_\alpha=\sum_{v\in V(\alpha)}\hat{V}_v\cdot
T_\alpha.\ea
Therefore, the total Hamiltonian constraint in this sector has been
quantized as a well-defined operator
$\hat{H}(N)=\sum_{i=1}^8\hat{H}_i$ in $\hil_\kin$. It is easy to see
that $\hat{H}(N)$ is internal gauge invariant and diffeomorphism
covariant. Hence it is at least well defined in the gauge invariant
Hilbert space $\hil_G$. However, it is difficult to define
$\hat{H}(N)$ directly on $\hil_{Diff}$. Moreover, as in $f(R)$
theories, the constraint algebra of STT do not form a Lie algebra. This
might lead to quantum anomaly after quantization.

\subsection{Sector of $\omega(\phi)= -3/2$ }

In the special case of $\omega(\phi)= -3/2$, the smeared version
$S(\lambda)$ of the extra
conformal constraint (\ref{conformalc}) has to be promoted as a well-defined operator. Note
that both $\phi$ and $\pi(R)$ are already well-defined operators. We
can use the following classical identity to quantize the conformal
constraint $S(\lambda)$,
\ba \kt\equiv\int_\Sigma
d^3x\tilde{K}^i_aE^a_i=\gamma^{-\frac32}\{\Euc(1),V\} \ea
where the
Euclidean scalar constraint $\Euc(1)$ by definition was: \ba
\Euc(1)&=&\frac{1}{2\kappa}\int_\Sigma
d^3xF^j_{ab}\frac{\varepsilon_{jkl} E^a_kE^b_l}{\sqrt{h}}. \ea Both
$\Euc$ and the volume $V$ have been quantized in LQG. Now it is easy
to promote $S(\lambda)$ as a well-defined operator, and its action
on a given basis vector $T_{\alpha,X}\in\hil_\kin$ reads \ba
\hat{S}(\lambda)\cdot T_{\alpha,X} &=& \left(\sum_{v\in
V(\alpha)}\frac{\lambda(v)}{\gamma^{3/2}\kappa(i\hbar)}[\hat{H}^E(1),\hat{V}_v]-\sum_{x\in
X}\lambda(x)\hat{\phi}(x)\hat{\pi}(x)\right)\cdot T_{\alpha,X}.\ea
It is clear that $\hat{S}(\lambda)$ is internal gauge invariant,
diffeomorphism covariant and graph-changing. Hence it is well
defined in $\hil_G$. The Hamiltonian constraint operator in this
sector is similar to that in the sector of $\omega(\phi)\neq -3/2$.
The difference is that now $\omega(\phi)= -3/2$. Hence we write Eq.
(\ref{hamilton1}) as $H(N)=\sum^5_{i=1}H_i$ in regular order. It is
easy to see that the terms $H_1,H_2,H_4,H_5$ just keep the same form
as those in last subsection, while the quantized version of $H_3$ is
\ba \hat{H}_3\cdot T_{\alpha,X} &=&-\sum_{v\in
V(\alpha)}\frac{2^{16}N(v)}{3^5\gamma^4(i\lambda_0)^2(i\hbar)^4\kappa E^2(v)}\hat{\phi}^{-1}(v)\nn\\
&\times&\sum_{v(\Delta)=v(\Delta')=v}\epsilon(s_L
s_M s_N)\epsilon^{LMN}\hat{U}^{-1}_{\lambda_0}(\phi(s_{s_L(\Delta)}))\nn\\
&\times&
[\hat{U}_{\lambda_0}(\phi(t_{s_L(\Delta)}))-\hat{U}_{\lambda_0}(\phi(s_{s_L(\Delta)}))]\nn\\
&\times&\Tr(\tau_i\hat{h}_{s_M(\Delta)}[\hat{h}^{-1}_{s_M(\Delta)},(\hat{V}_v)^{3/4}]
\hat{h}_{s_N(\Delta)}[\hat{h}^{-1}_{s_N(\Delta)},(\hat{V}_v)^{3/4}]) \nn\\
&\times&\epsilon(s_I
s_J s_K)\epsilon^{IJK}\hat{U}^{-1}_{\lambda_0}(\phi(s_{s_I(\Delta')}))\nn\\
&\times&
[\hat{U}_{\lambda_0}(\phi(t_{s_I(\Delta')}))-\hat{U}_{\lambda_0}(\phi(s_{s_I(\Delta')}))]\nn\\
&\times&\Tr(\tau_i\hat{h}_{s_J(\Delta')})[\hat{h}^{-1}_{s_J(\Delta')},(\hat{V}_{v})^{3/4}]
\hat{h}_{s_K(\Delta')}[\hat{h}^{-1}_{s_K(\Delta')},(\hat{V}_{v})^{3/4}])\cdot
T_{\alpha,X} . \ea Hence the total Hamiltonian constraint operator
$\hat{H}(N)=\sum^5_{i=1}\hat{H}_i$ now is also well defined in
$\hil_G$.

\section{master constraint\label{section5}}

In order to find the physical Hilbert space and avoid possible
quantum anomaly, master constraint programme was first introduced
into LQG by Thiemann in his seminal paper\cite{Th06}. The master constraint can be
employed to implement the Hamiltonian constraint. This programme
can also be generalized to the above quantum STT.

\subsection{Sector of $\omega(\phi)\neq -3/2$ }
In the sector of $\omega(\phi)\neq -3/2$, by definition, the
master constraint of the STT classically reads
\ba \mathcal
{M}:=\frac12\int_\Sigma d^3x\frac{\abs{H(x)}^2}{\sqrt{h}},
\label{mcs}\ea
where the expression of Hamiltonian constraint $H(x)$
is given by Eq. (\ref{hamilton}). The master constraint can be
regulated by a point-splitting strategy \cite{Ma061} as: \ba
\mathcal {M}^\epsilon=\frac12\int_\Sigma d^3y\int_\Sigma
d^3x\chi_\epsilon(x-y)\frac{{H(x)}}{\sqrt{V_{U^\epsilon_x}}}\frac{{H(y)}}{\sqrt{V_{U^\epsilon_y}}}.
\ea
Introducing a partition $\mathcal {P}$ of the 3-manifold
$\Sigma$ into cells $C$, we can get an operator
$\hat{H}^\varepsilon_{C,\beta}$ acting on the internal
gauge-invariant spin-scalar-network basis $T_{s,c}$ in $\hil_G$ via
a state-dependent triangulation, \ba
\hat{H}^\varepsilon_{C,\alpha}\cdot T_{s,c}=\sum_{v\in
V(\alpha)}\chi_C(v)\hat{H}^\varepsilon_v \cdot
T_{s,c}, \label{master1}\ea
where $\chi_C$ is the characteristic
function over $C$, $\alpha$ denotes the underlying graph of the
spin-network state $T_{s}$, and
\ba\hat{H}^\varepsilon_v=\sum_{v(\Delta)=v}\hat{H}^{\varepsilon,\Delta}_{GR,v}+\sum^8_{i=3}
\hat{H}^\varepsilon_{i,v}, \ea with \ba \hat{H}_{3,v}^\varepsilon
&=&
\frac{16\hat{f}(\phi(v))}{\gamma^3(i\hbar)^2\kappa}\hat{\phi}^{-1}(v)\nn\\
&\times&
[\hat{\Euc}(1),(\hat{V}_{U^\epsilon_{v}})^{1/4}][\hat{\Euc}(1),(\hat{V}_{U^\epsilon_{v}})^{1/4}],
\ea
\ba \hat{H}^\varepsilon_{4,v}
&=&-\sum_{v(\Delta)=v(\Delta')=v(X)=v}\frac{2^{18}\hat{f}(\phi(v))}{3^3\gamma^6(i\hbar)^6
E^2(v)}\hat{\pi}(v)
\nn\\
&\times&\Tr(\tau_i\hat{h}_{s_L(\Delta)}[\hat{h}^{-1}_{s_L(\Delta)},\hat{\kt}])\nn\\
&\times&\epsilon(s_L s_M s_N)\epsilon^{LMN}\Tr(\tau_i\hat{h}_{s_M(\Delta)}[\hat{h}^{-1}_{s_M(\Delta)},(\hat{V}_{U^\epsilon_{v}})^{1/2}]\nn\\
&\times&\hat{h}_{s_N(\Delta)}[\hat{h}^{-1}_{s_N(\Delta)},(\hat{V}_{U^\epsilon_{v}})^{1/2}]) \nn\\
&\times&\epsilon(s_I
s_J s_K)\epsilon^{IJK}\Tr(\hat{h}_{s_I(\Delta')}[\hat{h}^{-1}_{s_I(\Delta')},(\hat{V}_{U^\epsilon_{v}})^{1/2}]\nn\\
&\times&
\hat{h}_{s_J(\Delta')}[\hat{h}^{-1}_{s_J(\Delta')},(\hat{V}_{U^\epsilon_{v}})^{1/2}] \nn\\
&\times&\hat{h}_{s_K(\Delta')}[\hat{h}^{-1}_{s_K(\Delta')},(\hat{V}_{U^\epsilon_{v}})^{1/2}]),\ea
\ba \hat{H}^\varepsilon_{5,v}
&=&\sum_{v(\Delta)=v(\Delta')=v(X)=v}\frac{2^{20}\kappa\hat{f}(\phi(v))}{3^4\gamma^6(i\hbar)^6
E^2(v)}
\hat{\phi}(v)\hat{\pi}(v)\hat{\pi}(v) \nn\\
&\times&\epsilon(s_I s_J s_K)\epsilon^{IJK}\Tr(\hat{h}_{s_I(\Delta)}
[\hat{h}^{-1}_{s_I(\Delta)},(\hat{V}_{U^\epsilon_{v}})^{1/4}]\nn\\
&\times&
\hat{h}_{s_J(\Delta)}[\hat{h}^{-1}_{s_J(\Delta)},(\hat{V}_{U^\epsilon_{v}})^{1/2}] \nn\\
&\times&\hat{h}_{s_K(\Delta)}[\hat{h}^{-1}_{s_K(\Delta)},(\hat{V}_{U^\epsilon_{v}})^{1/2}]) \nn\\
&\times&\epsilon(s_L s_M s_N)\epsilon^{LMN}\Tr(\hat{h}_{s_L(\Delta')}[\hat{h}^{-1}_{s_L(\Delta')},(\hat{V}_{U^\epsilon_{v}})^{1/4}]\nn\\
&\times&
\hat{h}_{s_M(\Delta')}[\hat{h}^{-1}_{s_M(\Delta')},(\hat{V}_{U^\epsilon_{v}})^{1/2}] \nn\\
&\times&\hat{h}_{s_N(\Delta')}[\hat{h}^{-1}_{s_N(\Delta')},(\hat{V}_{U^\epsilon_{v}})^{1/2}]),
\ea
\ba \hat{H}^\varepsilon_{6,v}
&=&\sum_{v(\Delta)=v(\Delta')=v}\frac{2^{15}\hat{\omega}(\phi)}{3^4\gamma^4(i\lambda_0)^2(i\hbar)^4\kappa E^2(v)}\hat{\phi}^{-1}(v)\nn\\
&\times&\epsilon(s_L
s_M s_N)\epsilon^{LMN}\hat{U}^{-1}_{\lambda_0}(\phi(s_{s_L(\Delta)}))\nn\\
&\times&
[\hat{U}_{\lambda_0}(\phi(t_{s_L(\Delta)}))-\hat{U}_{\lambda_0}(\phi(s_{s_L(\Delta)}))]\nn\\
&\times&\Tr(\tau_i\hat{h}_{s_M(\Delta)}[\hat{h}^{-1}_{s_M(\Delta)},(\hat{V}_v)^{1/2}]
\hat{h}_{s_N(\Delta)}[\hat{h}^{-1}_{s_N(\Delta)},(\hat{V}_v)^{3/4}]) \nn\\
&\times&\epsilon(s_I
s_J s_K)\epsilon^{IJK}\hat{U}^{-1}_{\lambda_0}(\phi(s_{s_I(\Delta')}))\nn\\
&\times&
[\hat{U}_{\lambda_0}(\phi(t_{s_I(\Delta')}))-\hat{U}_{\lambda_0}(\phi(s_{s_I(\Delta')}))]\nn\\
&\times&\Tr(\tau_i\hat{h}_{s_J(\Delta')}[\hat{h}^{-1}_{s_J(\Delta')},(\hat{V}_{v})^{1/2}]
\hat{h}_{s_K(\Delta')}[\hat{h}^{-1}_{s_K(\Delta')},(\hat{V}_{v})^{3/4}]),
\ea
\ba \hat{H}^\varepsilon_{7,v}
&=&\frac{2^{9}}{3\gamma^2i\lambda_0(i\hbar)^2\kappa E(v)}\nn\\
&\times&\sum_{e(0)=v}X^i_e\sum_{v(\Delta)=v}\nn\\
&\times&\epsilon(s_I s_J s_K)\epsilon^{IJK}\hat{U}^{-1}_{\lambda_0}(\phi(s_{s_I(\Delta)}))\nn\\
&\times&
[\hat{U}_{\lambda_0}(\phi(t_{s_I(\Delta)}))-\hat{U}_{\lambda_0}(\phi(s_{s_I(\Delta)}))]\nn\\
&\times&\Tr(\tau_i\hat{h}_{s_J(\Delta)}[\hat{h}^{-1}_{s_J(\Delta)},(\hat{V}_{U^\epsilon_{v}})^{1/4}]\nn\\
&\times&
\hat{h}_{s_K(\Delta)}[\hat{h}^{-1}_{s_K(\Delta)},(\hat{V}_{U^\epsilon_{v}})^{1/4}]),
\ea
\ba \hat{H}^\varepsilon_{8,v}
&=&\frac1\kappa\hat{\xi}(\phi(v))\sqrt{\hat{V}_{U^\epsilon_{v}}};
\ea
Here $\hat{H}^{\varepsilon,\Delta}_{GR,v}$ keeps the same form
as the corresponding terms in \cite{Zh11b}. Note that the family of
operators $\hat{H}^\varepsilon_{C,\alpha}$ are cylindrically
consistent up to diffeomorphism. Thus the inductive limit operator
$\hat{H}_{C}$ is densely defined in $\hil_G$ by the uniform Rovelli-Smolin topology. Hence we could define master constraint operator
$\hat{\mathcal {M}}$ acting on a diffeomorphism invariant state as
\ba( \hat{\mathcal {M}}\Phi_{Diff})T_{s,c}=\lim_{\mathcal
{P}\rightarrow\Sigma,\varepsilon,\varepsilon'\rightarrow
0}\Phi_{Diff}[\frac12\sum_{c\in\mathcal
{P}}\hat{H}^\varepsilon_C(\hat{H}^{\varepsilon'}_C)^\dagger T_{s,c}
]. \ea Note that although the quantitative actions are different,
our construction of $\hat{\mathcal {M}}$ is qualitatively similar to
those in \cite{Zh11b,Ma06}. Similar methods in \cite{Ma06,Zh11b} can
be used here to prove that $\hat{\mathcal {M}}$ is a positive and
symmetric operator in $\hil_{Diff}$. Hence it admits a unique
self-adjoint Friedrichs extension. It is then possible obtaining the
physical Hilbert space of the quantum STT of this sector by the
direct integral decomposition of $\hil_{Diff}$ with respect to
$\hat{\mathcal {M}}$.

\subsection{Sector of $\omega(\phi)= -3/2$ }
In the case of $\omega(\phi)= -3/2$, both the implementation of the
Hamiltonian constraint and the implementation of the conformal constraint need to employ the
master constraint programme. We then define the master constraint for this
sector as
\ba \mathcal {M}:=\frac12\int_\Sigma
d^3x\frac{\abs{H(x)}^2+\abs{S(x)}^2}{\sqrt{h}}, \label{mcs1}\ea
where the expressions of Hamiltonian constraint $H(x)$ and the
conformal constraint $S(x)$ are given by Eqs. (\ref{hamilton1}) and
(\ref{conformalc}) respectively. It is clear that \ba \mathcal {M}=0
\Leftrightarrow H(N)=0 \quad and\quad S(\lambda)=0 \quad\forall
N(x),\lambda(x). \ea
Now the constraints form a Lie algebra. The
master constraint can be regulated by a point-splitting strategy as:
\ba \mathcal {M}^\epsilon=\frac12\int_\Sigma d^3y\int_\Sigma
d^3x\chi_\epsilon(x-y)\frac{H(x)H(y)+S(x)S(y)}{\sqrt{V_{U^\epsilon_x}}\sqrt{V_{U^\epsilon_y}}}.
\ea
Introducing a partition $\mathcal {P}$ of the 3-manifold
$\Sigma$ into cells $C$, we get an operator
$\hat{H}^\varepsilon_{C,\beta}$ acting on spin-scalar-network basis
$T_{s,c}$ in $\hil_G$ by a state-dependent triangulation as Eq.
(\ref{master1}). Here, note that $\hat{H}^\varepsilon_v$ has less
terms than in Eq. (\ref{master1}) as
\ba\hat{H}^\varepsilon_v=\sum_{v(\Delta)=v}\hat{H}^{\varepsilon,\Delta}_{GR,v}+\sum^5_{i=3}
\hat{H}^\varepsilon_{i,v}, \ea where \ba \hat{H}^\varepsilon_{3,v}
&=&-\sum_{v(\Delta)=v(\Delta')=v}\frac{2^{14}}{3^3\gamma^4(i\lambda_0)^2(i\hbar)^4\kappa E^2(v)}\hat{\phi}^{-1}(v)\nn\\
&\times&\epsilon(s_L
s_M s_N)\epsilon^{LMN}\hat{U}^{-1}_{\lambda_0}(\phi(s_{s_L(\Delta)}))\nn\\
&\times&
[\hat{U}_{\lambda_0}(\phi(t_{s_L(\Delta)}))-\hat{U}_{\lambda_0}(\phi(s_{s_L(\Delta)}))]\nn\\
&\times&\Tr(\tau_i\hat{h}_{s_M(\Delta)}[\hat{h}^{-1}_{s_M(\Delta)},(\hat{V}_v)^{1/2}]
\hat{h}_{s_N(\Delta)}[\hat{h}^{-1}_{s_N(\Delta)},(\hat{V}_v)^{3/4}]) \nn\\
&\times&\epsilon(s_I
s_J s_K)\epsilon^{IJK}\hat{U}^{-1}_{\lambda_0}(\phi(s_{s_I(\Delta')}))\nn\\
&\times&
[\hat{U}_{\lambda_0}(\phi(t_{s_I(\Delta')}))-\hat{U}_{\lambda_0}(\phi(s_{s_I(\Delta')}))]\nn\\
&\times&\Tr(\tau_i\hat{h}_{s_J(\Delta')}[\hat{h}^{-1}_{s_J(\Delta')},(\hat{V}_{v})^{1/2}]
\hat{h}_{s_K(\Delta')}[\hat{h}^{-1}_{s_K(\Delta')},(\hat{V}_{v})^{3/4}]),
\ea
and $H^\varepsilon_{4,v}$ and $H^\varepsilon_{5,v}$ keep the
same form as the corresponding terms in the sector of $\omega(\phi)\neq
-3/2$. The operator corresponding to the conformal constraint can be
defined in a similar way,
\ba \hat{S}^\varepsilon_{C,\alpha}\cdot
T_{s,c}=\sum_{v\in V(\alpha)}\chi_C(v)\hat{S}^\varepsilon_v \cdot
T_{s,c,}\ea
where
\ba\hat{S}^\varepsilon_v=\hat{S}^\varepsilon_{1,v}+\hat{S}^\varepsilon_{2,v},
\ea with \ba \hat{S}_{1,v}^\varepsilon &=&
\frac{2}{\gamma^{3/2}\kappa(i\hbar)}
[\hat{H}^E(1),(\hat{V}_{U^\epsilon_{v}})^{1/2}], \ea
\ba \hat{S}^\varepsilon_{2,v} &=&-\sum_{v(\Delta)=v(X)=v}\frac{
2^{7}}{3\gamma^3(i\hbar)^3E(v)}
\hat{\phi}(v)\hat{\pi}(v) \nn\\
&\times&\epsilon(s_I s_J s_K)\epsilon^{IJK}\Tr(\hat{h}_{s_I(\Delta)}
[\hat{h}^{-1}_{s_I(\Delta)},(\hat{V}_{U^\epsilon_{v}})^{1/2}]\nn\\
&\times&
\hat{h}_{s_J(\Delta)}[\hat{h}^{-1}_{s_J(\Delta)},(\hat{V}_{U^\epsilon_{v}})^{1/2}] \nn\\
&\times&\hat{h}_{s_K(\Delta)}[\hat{h}^{-1}_{s_K(\Delta)},(\hat{V}_{U^\epsilon_{v}})^{1/2}]).
\ea
Note that the family of operators
$\hat{H}^\varepsilon_{C,\alpha}$ and
$\hat{S}^\varepsilon_{C,\alpha}$ are cylindrically consistent up to
diffeomorphism. Hence the inductive limit operator $\hat{H}_{C}$ and
$\hat{S}_{C}$ can be densely defined in $\hil_G$ by the uniform
Rovelli- Smolin topology. Thus we could define master constraint
operator $\hat{\mathcal {M}}$ acting on diffeomorphism invariant
states as \ba( \hat{\mathcal {M}}\Phi_{Diff})T_{s,c}=\lim_{\mathcal
{P}\rightarrow\Sigma,\varepsilon,\varepsilon'\rightarrow
0}\Phi_{Diff}[\frac12\sum_{c\in\mathcal
{P}}\left(\hat{H}^\varepsilon_C(\hat{H}^{\varepsilon'}_C)^\dagger+\hat{S}^\varepsilon_C(\hat{S}^{\varepsilon'}_C)^\dagger\right)
T_{s,c} ]. \ea Similarly, we can prove that $\hat{\mathcal {M}}$ is
a positive and symmetric operator in $\hil_{Diff}$ and hence admits
a unique self-adjoint Friedrichs extension\cite{Ma06,Zh11b}. Hence
it is also possible to obtain the physical Hilbert space of the
quantum STT in this special case by the direct integral
decomposition of $\hil_{Diff}$ with respect to the spectrum of
$\hat{\mathcal {M}}$.

\section{Cosmological application of quantum Brans-Dikce theory \label{section6}}

For cosmological application of above loop quantum STT, in this section,
we will set up the basic structure of loop quantum Brans-Dicke cosmology and get its effective equations of motion \cite{ZM12b}. For simplicity, we only consider the spatially flat ($k=0$) homogeneous and isotropic universe in Brans-Dicke theory. Recall that the original Brans-Dicke theory is the particular case of STT
with constant $\omega$ and vanishing potential of $\phi$. Thus the Hamiltonian constraint of Brans-Dicke theory reads \ba
H&=&\frac{\phi}{2\kappa}\left[F^j_{ab}-(\gamma^2+\frac{1}{\phi^2})\varepsilon_{jmn}\tilde{K}^m_a\tilde{K}^n_b\right]\frac{\varepsilon_{jkl}
E^a_kE^b_l}{\sqrt{h}}\nn\\
&+&\frac{\kappa}{3+2\omega}\left(\frac{(\tilde{K}^i_aE^a_i)^2}{\kappa^2\phi\sqrt{h}}+
2\frac{(\tilde{K}^i_aE^a_i)\pi}{\kappa\sqrt{h}}+\frac{\pi^2\phi}{\sqrt{h}}\right) \nn\\
&+&\frac{\omega}{2\kappa\phi}\sqrt{h}(D_a\phi)
D^a\phi+\frac{1}{\kappa}\sqrt{h}D_aD^a\phi.\label{BDhamilton} \ea
In the cosmological model, classically the metric of spacetime can be written as the
following Friedman-Robertson-
Walker (FRW) formalism,
\ba
ds^2=-dt^2+a^2(t)[dr^2+r^2(d\theta^2+\sin^2\theta d\phi^2)] \ea
where $a$ is the scale factor.
The classical Friedman equation
of Brans-Dicke cosmology reads
\ba \left(\frac{\dot{a}}{a}
+\frac{\dot{\phi}}{2\phi}\right)^2&=&\frac{2\omega+3}{12}\left(\frac{\dot{\phi}}{\phi}\right)^2+\frac{8\pi
G\rho}{3\phi}.\ea
Our task is to quantize this model by the loop quantization method and find the quantum dynamical equation as well as its effective expression.

\subsection{Loop quantum Brans-Dicke cosmology}

Since the space of our cosmological model is infinite, we introduce an ``elemental
cell" $\mathcal {V}$ and restrict all
integral to $\mathcal {V}$. The homogeneity of the universe guarantee that the whole space
information is reflected in this elemental cell. Now we choose a fiducial
Euclidean metric  $ {}^oq_{ab}$ and introduce a pair of fiducial orthnormal
triad and co-triad as $({}^oe^a_i , {}^o\omega^i_a)$ respectively
such that $ {}^oq_{ab}={}^o\omega^i_a{}^o\omega^i_b$. For simplicity,
we let the elemental cell $\mathcal {V}$ be a cubic measured by our
fiducial metric and denotes its volume as $V_o$. Because our FRW metric
is spatially flat, we have $\Gamma_a^i=0$ and hence
$A_a^i=\gamma \kt_a^i$. Via fixing the degrees of freedom of local
gauge and diffeomorphism, we finally obtain the connection and
densitized triad by symmetrical reduction as \cite{LQC5}:
\ba A_a^i=\ct
V_0^{-\frac13}{}^o\omega^i_a,\quad\quad\quad
E^b_j=pV_0^{-\frac23}\sqrt{\det({}^0q)}{}^oe^b_j, \ea
where $\ct,p$
are only functions of $t$.  Hence the phase space of the cosmological model consists of conjugate pairs $(\ct,p)$ and
$(\phi,\pi)$. The basic Poisson brackets between them can be simply
read as\ba
\{\ct,p\}&=&\frac{\kappa}{3}\gamma,\nn\\
\{\phi,\pi\}&=&1. \label{poissonb}\ea
Note that by the symmetric reduction, the Gaussian and
diffeomorphism constraints are satisfied automatically. Also
in our homogeneous model, the
last two spatial derivative terms in the Hamiltonian constraint (\ref{BDhamilton}) can be neglected. Hence we only need to consider the first five terms in (\ref{BDhamilton}). The reduced
Hamiltonian in the cosmological model reads \ba
H&=&-\frac{3\ct^2\sqrt{\abs{p}}}{\gamma^2\kappa\phi}+\frac{\kappa}{(3+2\omega)\phi
\abs{p}^{\frac32}}(\frac{3\ct p}{\kappa\gamma}+\pi\phi)^2.\ea

To quantize the cosmological model, we first need to construct the
quantum kinematic of Brans-Dicke cosmology by mimicking the loop
quantum STT. This is the so-called polymer-like quantization. The
kinematic Hilbert space for the geometry part can be defined as
$\mathcal{H}_{\kin}^{\grav}:=L^2(R_{Bohr},d\mu_{H})$, where
$R_{Bohr}$ and $d\mu_{H}$ are respectively the Bohr
compactification of the real line and Haar measure on it
\cite{LQC5}. On the other hand, for convenience we choose
Schrodinger representation for the scalar field \cite{AS11}. Thus
the kinematic Hilbert space for the scalar field part is defined as
in usual quantum mechanics,
$\mathcal{H}_{\kin}^{\sca}:=L^2(R,d\mu)$. Hence the whole Hilbert
space is a direct product, $\hil_\kin^{total}=\hil^\grav_\kin\otimes
\hil^\sca_\kin$. Now let $\ket{\mu}$ be the eigenstates of
 $\hat{p}$ in the kinematic Hilbert space $\mathcal{H}_{\kin}^{\grav}$ such that
 \ba
\hat{p}\ket{\mu}=\frac{8\pi G\gamma\hbar}{6}\mu\ket{\mu}. \ea
It turns out that those
states satisfy the following orthonormal condition
\ba
\bra{\mu_i}{\mu_j}\rangle=\delta_{\mu_i,\mu_j}\ , \ea
where $\delta_{\mu_i,\mu_j}$ is the Kronecker delta function rather than the Dirac distribution.
For the convenience of studying quantum dynamics, we define new
variables
 \ba
v:=2\sqrt{3}\sgn(p)\mubar^{-3},\quad b:=\mubar \ct, \ea
where $\sgn(p)$ is the sign function for
$p$ and
$\mubar=\sqrt{\frac{\Delta}{\abs{p}}}$ with
$\Delta=4\sqrt{3}\pi\gamma{\ell}_{\textrm{p}}^2$ being a minimum
nonzero eigenvalue of the area operator \cite{Ash-view}. They also
form a pair of conjugate variables as
\ba \{b,v\}=\frac{2}{\hbar}.
\ea
It turns out that the eigenstates of
 $\hat{v}$ also contribute an orthonormal basis in $\mathcal{H}_{\kin}^{\grav}$.
We denote
$\ket{\phi,v}$ as the orthogonal basis for the whole Hilbert space
$\hil^{total}_\kin$.

Now we come to the quantum dynamics. We treat the first two terms
of Hamiltonian constraint (\ref{BDhamilton}) in the same way as in standard LQC \cite{aps1}. Hence, the first
two terms of the Hamiltonian constraint act on a quantum state
$\Psi(\nu,\phi)\in\hil^{total}_\kin$ as
\ba
(\hat{H}_1+\hat{H}_2)\Psi(\nu,\phi)=\frac1\phi\left(f_+(v)\Psi(\nu+4,\phi)+f_0(v)\Psi(\nu,\phi)+f_-(v)\Psi(\nu-4,\phi)\right),
\ea
where
\ba
f_+(v)=\frac{\sqrt{3\Delta}}{16\kappa\gamma^2}\Abs{\abs{v+3}-\abs{v+1}}\abs{v+2},\nn\\
f_-(v)=f_+(v-4), \quad  f_0(v)=-f_+(v)-f_-(v).\ea Then we turn to
the third term $H_3\equiv
\frac{\kappa}{3+2\omega}\frac{(\tilde{K}^i_aE^a_i)^2}{\kappa^2\phi\sqrt{h}}$.
Due to spatial flatness, we have
$\kt_a^iE^a_i=\frac{1}{\gamma}A_a^iE^a_i$. In the cosmological
model, this term can be reduced by
\ba
\frac{1}{\gamma}A_a^iE^a_i\rightsquigarrow \frac{3}{\gamma}\ct
p=\frac{3\kappa\hbar bv}{4}. \ea
Because we use polymer
representation for geometry, their is no quantum operator
corresponding to connection $\ct$ as in standard LQC \cite{AS11}.
Hence we have to replace the connection by holonomy to get a
well-defined operator. It turns out that the term $H_3$ can be
quantized, and its action on a quantum state reads \cite{ZM12b}
\ba
\hat{H}_3\Psi(\phi,v)&=&\frac{2\sqrt{3}\kappa}{\beta\phi(\Delta)^{\frac32}}\left(\frac{3\hbar
}{4}\right)^2
\sin(b)\hat{|v|}\sin(b)\Psi(\phi,v)\nn\\
&=&-\frac{\sqrt{3}\kappa}{2\beta\phi(\Delta)^{\frac32}}\left(\frac{3\hbar
}{4}\right)^2\left[\abs{v+2}\Psi(\phi,v+4)-2\abs{v}\Psi(\phi,v)+\abs{v-2}\Psi(\phi,v-4)\right],\ea
where we set $\beta=3+2\omega$. Similarly, the fourth term
$H_4\equiv \frac{2\kappa}{3+2\omega}
\frac{(\tilde{K}^i_aE^a_i)\pi}{\kappa\sqrt{h}}$ can also be
quantized, and its action on a wave function reads
\ba
\hat{H}_4\Psi(\phi,v)&=&\frac{2\sqrt{3}\kappa}{\beta(\Delta)^{\frac32}}\left(\frac{3\hbar
}{4}\right)2\sgn(p)\sin(b)\hat{\pi}\Psi(\phi,v)\nn\\
&=&\frac{2\sqrt{3}\kappa}{\beta(\Delta)^{\frac32}}\left(\frac{3\hbar
}{4}\right) \hbar\sgn(p)[\frac{\partial\Psi(\phi,v+2)}{\partial\phi}
-\frac{\partial\Psi(\phi,v-2)}{\partial\phi}]. \ea
The last term
$H_5\equiv \frac{\kappa}{3+2\omega}\frac{\pi^2\phi}{\sqrt{h}}$ can be quantized as
\ba
\hat{H}_5\Psi(\phi,v)&=&\frac{2\sqrt{3}\kappa}{\beta(\Delta)^{\frac32}}
\widehat{|v|^{-1}}(\hat{\pi})\hat{\phi}\hat{\pi}\Psi(\phi,v)\nn\\
&=&-\frac{2\sqrt{3}\kappa}{\beta(\Delta)^{\frac32}}(\hbar)^2B(v)\phi\frac{\partial^2\Psi(\phi,v)}{\partial\phi^2},\ea
where \ba
B(v)=(\frac32)^3\abs{v}\abs{\abs{v+1}^{1/3}-\abs{v-1}^{1/3}}^3. \ea
The total Hamiltonian constraint equation of loop quantum Brans-Dicke cosmology reads
\ba
(\sum^5_{i=1}\hat{H}_i)\Psi(\phi,v)=0.\label{hbd}
\ea

\subsection{Effective equation}

To study the effective theory of loop quantum Brans-Dicke cosmology, we also want to know the effect of matter fields on the dynamical evolution. Hence we include an extra massless scalar matter field $\varphi$ into Brans-Dicke cosmology.
Then classically the total Hamiltonian constraint of the model
reads
\ba
H=-\frac{3\ct^2\sqrt{\abs{p}}}{\gamma^2\kappa\phi}+\frac{\kappa}{(3+2\omega)\phi
\abs{p}^{\frac32}}(\frac{3\ct
p}{\kappa\gamma}+\pi\phi)^2+\frac{p_\varphi^2}{2\abs{p}^{\frac32}}, \label{Hcouple}\ea
where $p_\varphi$ is the momentum conjugate to $\varphi$. The effective description of LQC is a delicate and topical issue since it may relate the
quantum gravity effects to low-energy physics. The effective equations of LQC are being studied from both canonical
perspective\cite{Taveras,DMY,YDM,Boj11} and path integral perspective\cite{ACH102,QHM,QDM,QM1,QM2}. Since the key element in
the polymer-like quantization of previous subsection is to take holonomies rather than connections as basic variables, a heuristic and simple way to
get the
effective equations is to do the replacement $\ct\rightarrow
\frac{\sin(\mubar\ct)}{\mubar}$ or $b\rightarrow
\sin b$. Under this replacement, the effective version of
Hamiltonian constraint (\ref{Hcouple}) takes the form
\ba H=-\frac{3\sin^2(\mubar\ct)\sqrt{\abs{p}}}{\kappa\gamma^2\phi\mubar^2}+\frac{\kappa}{\beta\phi
\abs{p}^{\frac32}}(\frac{3\sin(\mubar\ct)
p}{\mubar\kappa\gamma}+\pi\phi)^2+\abs{p}^{\frac32}\rho, \label{Heffective}\ea
where $\rho=\frac{p_\varphi^2}{2\abs{p}^3}$  by definition is the
matter density. It is worth noting that the effective Hamiltonian (\ref{Heffective}) can also be derived by a path integral formalism \cite{ZM12b}. Then the canonical equations of motion
read
\ba
\dot{p}&=&\frac{2\sqrt{\abs{p}}}{\gamma\phi\mubar}\sin(\mubar\ct)\cos(\mubar\ct)-\frac{2\kappa}{\beta\phi
\abs{p}^{\frac12}}\sgn(p)\left(\frac{3\sin(\mubar\ct)
p}{\mubar\kappa\gamma}+\pi\phi)\cos(\mubar\ct\right),\label{pdot}\\
\dot{\phi}&=& \frac{2\kappa}{\beta
\abs{p}^{\frac32}}(\frac{3\sin(\mubar\ct)
p}{\mubar\kappa\gamma}+\pi\phi)\label{phidot}.\ea
In the above
calculation, the Poisson brackets (\ref{poissonb}) were used.
The Combination
of equations (\ref{pdot}) and (\ref{phidot}) gives
\ba
\left(\frac{\dot{p}}{2p}+\frac{\dot{\phi}}{2\phi}\right)^2&=&
\left[\frac{\sgn(p)}{\gamma\phi\mubar\sqrt{\abs{p}}}\sin(\mubar\ct)\cos(\mubar\ct)+\frac{\kappa}{\beta\phi
\abs{p}^{\frac32}}(\frac{3\sin(\mubar\ct)
p}{\mubar\kappa\gamma}+\pi\phi)(1-\cos(\mubar\ct))\right]^2\nn\\
&=&\left[\frac{\sgn(p)}{\gamma\phi\sqrt{\Delta}}\sin(\mubar\ct)\cos(\mubar\ct)+\frac{\dot{\phi}}{2\phi
}(1-\cos(\mubar\ct))\right]^2. \label{hubble}\ea
On the other hand,
from effective Hamiltonian constraint (\ref{Heffective}) we can get
\ba
-\frac{3\sin^2(\mubar\ct)}{\kappa\gamma^2\phi\Delta}+\frac{\beta\dot{\phi}^2}{4\kappa\phi
}+\rho=0, \ea
which implies
\ba
\sin^2(\mubar\ct)=\frac{\rho_\eff}{\rho_c}, \label{sin}\ea
where
$\rho_c=\frac{3}{\gamma^2\Delta\kappa}=\frac{\sqrt{3}}{32\pi^2G^2\gamma^3\hbar}$
and $\rho_\eff=\frac{\beta\dot{\phi}^2}{4\kappa}+\phi\rho$. Taking account of Eq. (\ref{sin}), we can rewrite Eq. (\ref{hubble}) as
\ba
\left(\frac{\dot{a}}{a}+\frac{\dot{\phi}}{2\phi}\right)^2
&=&\left[\frac{1}{\phi}\sqrt{\frac{\kappa}{3}\rho_\eff(1-\frac{\rho_\eff}{\rho_c})}+\frac{\dot{\phi}}{2\phi
}(1-\sqrt{1-\frac{\rho_\eff}{\rho_c}})\right]^2.  \label{tildeH}\ea
This is the effective Friedmann equation of Brans-Dicke cosmology, which contains important quantum correction terms. In addition, we can show that for a contracting universe, $\rho_\eff$ monotonically increase while $v$ decreases\cite{ZM12b}. Thus it is easy to see from Eq.(\ref{tildeH}) that, when $\rho_\eff$ approaches $\rho_c$, one gets
$\cos(\mubar\ct)=1-\frac{\rho_\eff}{\rho_c}=0$. Then from Eq.
(\ref{pdot}), we can obtain $\dot{p}=0$. This implies a quantum bounce would
happen at that point for a contracting universe.

We end
up this section with several remarks. First,
when $\phi=1$, because of $\rho_\eff=\rho$ and $\dot{\phi}=0$
we would return to the well-known effective Friedmann equation of LQC \cite{aps1,DMY} as
\ba
\left(\frac{\dot{a}}{a}\right)^2
&=&\frac{\kappa}{3}\rho(1-\frac{\rho}{\rho_c}). \ea
Second, when $\rho_\eff\ll\rho_c$, we can omit
$\frac{\rho_\eff}{\rho_c}$ terms in Eq. (\ref{tildeH}) to get the
classical limit of this equation as
\ba
\left(\frac{\dot{a}}{a}+\frac{\dot{\phi}}{2\phi}\right)^2
&=&\frac{1}{\phi^2}\frac{\kappa}{3}\rho_\eff=
\frac{\kappa}{3\phi^2}(\frac{\beta\dot{\phi}^2}{4\kappa}+\phi\rho)\nn\\
&=&\frac{\beta\dot{\phi}^2}{12\phi^2}+\frac{\kappa\rho}{3\phi},\ea
which is nothing but the classical Friedmann equation of Brans-Dicke
cosmology. Hence the effective theory has correct classical limit.

\section{Conclusion and outlook\label{section7}}

Modified gravity has received increased attention in issues of
``dark matter", ``dark energy" and nontrivial tests on gravity
beyond GR. Some kinds of modified gravity theories have also become
popular in certain unification schemes such as string theory. Whether some modified gravity theories could be nonperturbatively quantized is certainly an interesting and challenging question. In this review, as an example, we first set up
the $su(2)$-connection dynamical formalism of STT. Then LQG method has
been successfully extended to the STT by coupling to a polymer-like
scalar field. This successful extension strongly hints that the nonperturbative quantization procedure
might be valid even for more general modified gravity theories. At least, as we demonstrated, loop quantization procedure should be valid for any
metric theories with a well-defined geometrical dynamics. Hence it is
desirable to study the Hamiltonian formulation of modified gravity theories
and try to cast those theories into the $su(2)$-connection dynamical
formalism. Then, we can naturally extend nonperturbative loop
quantization method to those theories.

The concrete results of this paper are summarized as follows. A general loop quantization scheme for metric modified gravity is first given in
section \ref{section2}. Then we use STT as an example to show how
our general procedure works. By doing Hamiltonian analysis, we have
successfully derived the Hamiltonian formulation of STT of gravity
from their Lagrangian formulation. The result shows that these
theories can be naturally divided into two different sectors by the coupling parameter
$\omega(\phi)$. In the first sector of $\omega(\phi)\neq -3/2$, the
resulted canonical structure and constraint algebra of STT are
similar to those of GR minimally coupled with a scalar field. While
in the sector of $\omega(\phi)= -3/2$, the feasible theories are
strongly restricted and a new primary constraint which generating
conformal transformations of spacetime is obtained. The
corresponding canonical structure and constraint algebra are also
obtained. It is worth noting that the Palatini $f(\R)$ theories are
equivalent to this sector of STT. The successful background
independent LQG relies on the key observation that GR can be cast
into the $su(2)$-connection dynamics. We have shown that the
connection dynamical formalism of the STT can also be obtained by
canonical transformations from the geometrical dynamics. Based on the
connection dynamical formalism with structure group $SU(2)$, loop
quantization method has been successfully extended to the STT by
coupling to a polymer-like scalar field. The quantum kinematical
structure of STT is as same as that of loop quantum gravity coupled
with a scalar field. Thus the important physical result that both
the area and the volume are discrete at kinematic level remains
valid for quantum STT of gravity. While the dynamics of STT is more
general than that of LQG, the Hamiltonian constraint operators and
master constraint operators for STT can also be well defined in both
sectors respectively. In particular, in the sector $\omega(\phi)=
-3/2$, the extra conformal constraint can also be promoted as a
well-defined operator. Hence the classical STT in both sectors have been
successfully quantized non-perturbatively. This ensures the existence
of the STT of gravity at fundamental quantum level. As the cosmological application of the above loop quantum STT, we construct a particular type of loop quantum scalar-tensor
cosmology, which is the so-called Brans-Dicke cosmology. For
simplicity, we only restrict ourselves to the sector of $\omega\neq-\frac32$. It turns out that
the classical differential equation of Brans-Dicke cosmology, which represents the
cosmological evolution, is now replaced by quantum difference
equation. The effective Friedmann equation of loop quantum Brans-Dicke
cosmology is also given, which shows that the classical big bang singularity is again
replaced by a quantum bounce. This effective equation lays a foundation for further phenomenological investigation to possible quantum gravity effects in Brans-Dicke cosmology.

It should be noted that there are still many aspects of the
connection formalism and loop quantization of modified gravity, which
deserve discovering. Taking STT for examples, it is still desirable to
derive the connection dynamics of STT by variational principle. The
semiclassical analysis of loop quantum STT is yet to be done. In our loop quantum Brans-Dicke cosmology, some phenomenological issues, such as inflation, would be studied in future works. To
further explore the physical contents of the loop quantum STT, we
would also like to study its applications to black holes in
future works. In addition, one would also like to quantize STT via
the covariant spin foam approach. Furthermore, nonperturbative
loop quantization of other types of modified gravity, such as
Horava-Lifshitz theory and critical gravity etc, is also desirable.

\begin{acknowledgements}

This work is supported by NSFC (Grant No.10975017, No.11235003 and No.11275073)
and the Fundamental Research Funds for the Central Universities
under Grant No.2012ZZ0079.
\end{acknowledgements}

\end{document}